\documentclass[
twocolumn,
 amsmath,amssymb,
 aps,
pra,
 showpacs
]{revtex4-1}
\usepackage{placeins}
\usepackage{amsmath}
\usepackage{mathrsfs}
\usepackage{graphicx}
\usepackage{dcolumn}
\usepackage{bm}
\usepackage{ulem}
 \usepackage{relsize}
\usepackage[
margin=.7in,
]{geometry}

\begin{document}
\title{Understanding molecular harmonic emission at relatively long intense laser pulses: Beyond the Born-Oppenheimer approximation
}

\author{H. Ahmadi$^{1}$ }
\author{M. Vafaee$^{2}$}
\author{A. Maghari$^{1}$}

\affiliation{
$^{1}$Department of Physical Chemistry, School of Chemistry, College of Science, University of Tehran, Tehran, Iran
\\$^{2}$Department of Chemistry, Tarbiat Modares University, P. O. Box 14115-175, Tehran, Iran
 }

\begin{abstract}
The underlying physics behind the molecular harmonic emission in relatively long sin$^2$-like laser pulses is investigated.
We numerically solved  the full-dimensional electronic time-dependent Schr\"{o}dinger equation beyond the Born-Oppenheimer approximation for  simple molecular ion H$_2^+$. The occurrence and the effect of electron localization, non-adiabatic redshift and spatially asymmetric emission are evaluated to understand better complex patterns appearing in the high-order harmonic generation (HHG) spectrum. 
 Results show that the complex patterns in the HHG spectrum  originate mainly from a non-adiabatic response of the molecule to the rapidly changing laser field and also from a spatially asymmetric emission along the polarization direction. The effect of electron localization on the HHG spectrum was not observed as opposed to what is reported in the literature.
\end{abstract}

\maketitle
\section{Introduction}
Femtosecond and sub-femtosecond laser pulses with well-defined electric field shapes have enabled 
 us to steer and control nuclear and electronic dynamics in their natural time scales. The non-perturbative interaction of such laser pulses with molecules leads to different phenomena, such as high-order harmonic generation (HHG), above-threshold ionization and dissociation, bond hardening and softening [1]. 
 
 Among the above-mentioned phenomena, HHG has been attracted great attention during the last decades. The mechanism of HHG is  well understood by a three-step model proposed by Corkum [2] and extended by Lewenstein \textit{et al}. [3]. First, an electron tunnels into a continuum from a suppressed potential created with the combination of  the system's Coulomb potential and the laser field. Then, the released electron oscillates in the laser field and moves away from the ion core, and after a sign reversal of the field, it is driven back to the core. Finally, the ionized electron may recombine with its parent ion, leading to  radiation at multiples of the driving field's frequency. HHG is used to produce single isolated or trains of attosecond laser pulses, permitting real-time observation of electronic dynamics [4].
 
The emitted photons in the HHG process can also be analyzed to retrieve both structural and dynamical information of a medium [5-11]. 
For example, the time-dependent internuclear distance of a molecule can be retrieved from the frequency modulation observed in the HHG spectrum [11]. For Gaussian-like laser pulses having  rising and falling parts, the effective amplitude of each cycle experienced by a medium  changes non-adiabatically from a laser cycle to another. The non-adiabatic response of a medium to this rapidly changing laser field leads to a frequency modulation in the HHG process non-adiabatically such as a frequency blue-shift (red-shift) of the harmonics  at the rising (falling) part of a laser pulse [12-15].

 Another interesting phenomenon in intense laser fields is laser-induced localization of an electron on a specific proton in a dissociative molecular ion [16-17]. Based on quantum mechanics, a superposition of two states with different parity of a molecular ion leads to  electron localization on either one of the nuclei as internuclear separation grows.  The molecular ion is usually formed by the interaction of a femtosecond or an attosecond laser pulse with its parent molecule, launching a nuclear wavepacket in the ground state of the molecular ion. Being the simplest two- and single-electron molecules, molecular hydrogen H$_2$ and its ion H$_2^+$  have been considered as benchmarks for investigating the localization of an electron.  For example, in the first experiment on  electron localization in D$_2^+$,  the ion was formed from the ionization of the molecular D$_2$ by a 5 femtosecond laser pulse [18]. In order to prepare a superposition of states with a different parity, it is needed to excite hydrogen-like molecular ions from the ground electronic state $1s\sigma _g$  into the first excited state $2p\sigma _u$. This step can be carried out by the same pulse (single-pulse scheme), initially used for the ionization [18-20], or by a second laser pulse (two-pulse scheme) [21-22]. The observed asymmetry as a result of  electron localization depends strongly on the carrier-envelope phase of the driving laser pulse in a single-pulse scheme, and the time delay between two laser pulses in a two-pulse scheme. In general, the degree of electron localization in a two-pulse scheme is larger than that in a single-pulse one. As the driving pulse couples the two electronic states, the electron can be viewed as it is being transfered between the left and right nucleus. As the distance between the two nuclei increases (larger than 6 a.u.), the potential barrier between the two nuclei also rises. Finally, the electron wavepacket is trapped on either one of the nuclei and localization is frozen.
 The electron localization is experimentally characterized  by the asymmetry measured in a number of the emitted protons dissociating in the opposite directions along the internuclear axis. 

In this work, we seek underlying physics behind the harmonic emission in H$_2^+$ under relatively long sin$^2$-like laser pulses. Morales \textit{et al.} have recently reported for a one-dimensional  H$_2^+$ under linearly polarized 14-cycle sin$^2$ laser pulses that even-order harmonics  are produced  as a result of  electron localization [23]. They attributed the appearance of even-order harmonics to symmetry breaking of the system due to final electron localization  at relatively large internuclear separations.
We recently showed that the HHG spectrum gets complex due to the influence of a few-cycle pulse trailing edge [24,25]. These complicated patterns were attributed to the non-adiabatic redshift [24,25] and spatially asymmetric emission [25]. To our knowledge, a comprehensive study of a molecular HHG complexity in long sin$^2$- or Gaussian-like laser pulses beyond the Born-Oppenheimer approximation has not been  addressed.
Here, we consider electron localization, non-adiabatic redshift and spatially asymmetric emission to understand the complex patterns observed in sin$^2$-like laser pulses. 
 The HHG spectrum is analyzed  by calculating different parameters representing time-dependent electron localization for different simulations, with and without existence of considerable final electron localization. In addition, we decompose the total HHG spectrum into different localized signals to find out the origin of the observed complex patterns.

   To do so, the full-dimensional  electronic time-dependent Schr\"{o}dinger equation (TDSE) beyond the Born-Oppenheimer approximation (NBO) is numerically solved  for H$_2^+$. Calculations have been done with relatively long laser pulses with the Gaussian, Sin$^2$ and trapezoidal envelopes at 790 and 800 nm wavelengths and $I=$3 and 9 $\times 10^{14}$ Wcm$^{-2}$ intensities.  
 We assume that the molecular ion is aligned with its internuclear-distance axis parallel to the laser polarization direction. The molecular alignment  is readily implied experimentally nowadays [5,8,26-31]. We use atomic units throughout the article unless stated otherwise.

\section{Computational Methods}
We considered $z$ and $\rho$ as electron cylindrical coordinates, which are measured with respect to the center of mass of the two nuclei. The nuclear motion is described in the spherical coordinate with only variable $R$, which represents the internuclear distance. Therefore, we ignored the molecular rotation (with $\theta$ and $\varphi$ variables).
The time-dependent Schr\"{o}dinger equation for H$_2^+$  for both $z$ and $R$ parallel to the laser polarization direction, can be expressed (after separation of the center-of-mass motion) as [32-34]

\begin{eqnarray}\label{eq:1}
  i \frac{\partial \psi(z,\rho, R;t)}{\partial t}={\widehat H}(z,\rho, R;t)\psi(z,\rho, R;t).
\end{eqnarray}
 \^{H} is the total electronic and nuclear Hamiltonian which is given by
\begin{eqnarray}\label{eq:2}
 \widehat{H}(z,\rho, R;t)=&\mathlarger{-\frac{2m_N+m_e}{4m_Nm_e}[\frac{\partial^2}{\partial \rho^2}+\frac{1}{\rho}\frac{\partial}{\partial \rho}+\frac{\partial^2}{\partial z^2}]}
\nonumber \\
& \mathlarger{-\frac{1}{m_N}\frac{\partial^2}{\partial R^2}+V_C(z,\rho, R;t)},
\end{eqnarray}
with
\begin{eqnarray}\label{eq:3}
 \widehat{V}_C&(z,\rho, R,t)=\mathlarger{-\frac{1}{\sqrt{(z+\frac{R}{2})^2+\rho^2}}-\frac{1}{\sqrt{(z-\frac{R}{2})^2+\rho^2}}}
\nonumber \\
 &\mathlarger{+\frac{1}{R}+(\frac{2m_N+2m_e}{2m_N+me})zE_0f(t)cos(\omega t+\phi)}.
\end{eqnarray}
In these equations, $E_0$ is the laser peak amplitude, $m_e$ and $m_N$ are, respectively, the electron
and proton masses, $\omega$ is the angular frequency, $\phi$ is the carrier-envelope phase (CEP),  and \textit{f}(t) is the laser pulse envelope. For the trapezoidal pulse, the envelope  rises linearly during the first two cycles, then is constant for 10 cycles and decreases during the last two cycles.

\begin{figure*}[ht]
\begin{center}
\begin{tabular}{c}
\centering
\resizebox{175mm}{55mm}{\includegraphics{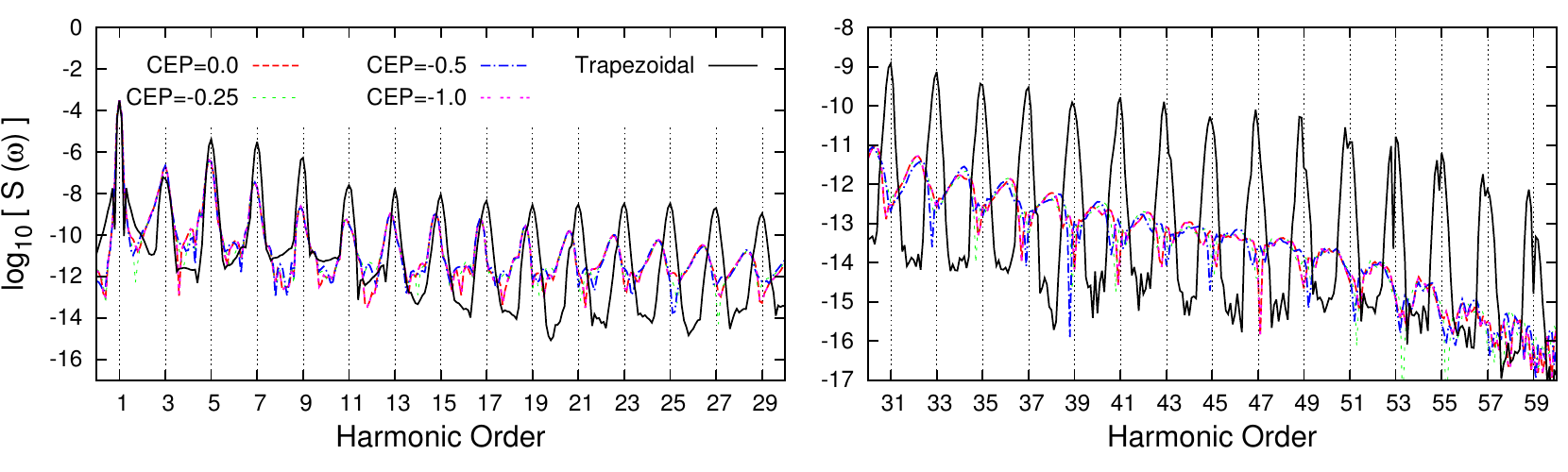}}
\end{tabular}
\caption{
\label{HHG} 
(Color online) High-order harmonic spectra produced by H$_2^+$ under 14-cycle sin$^2$ (calculated for different CEP values) and trapezoidal laser pulses of 800 nm wavelength at $I=$3 $\times 10^{14}$ Wcm$^{-2}$ intensity. For better clarity, the range of 1-29 and 31-59 harmonics of the spectra are shown in the left and right panels, respectively.		}
\end{center}
\end{figure*}
\begin{figure}[ht]
\centering
\resizebox{90mm}{130mm}{\includegraphics{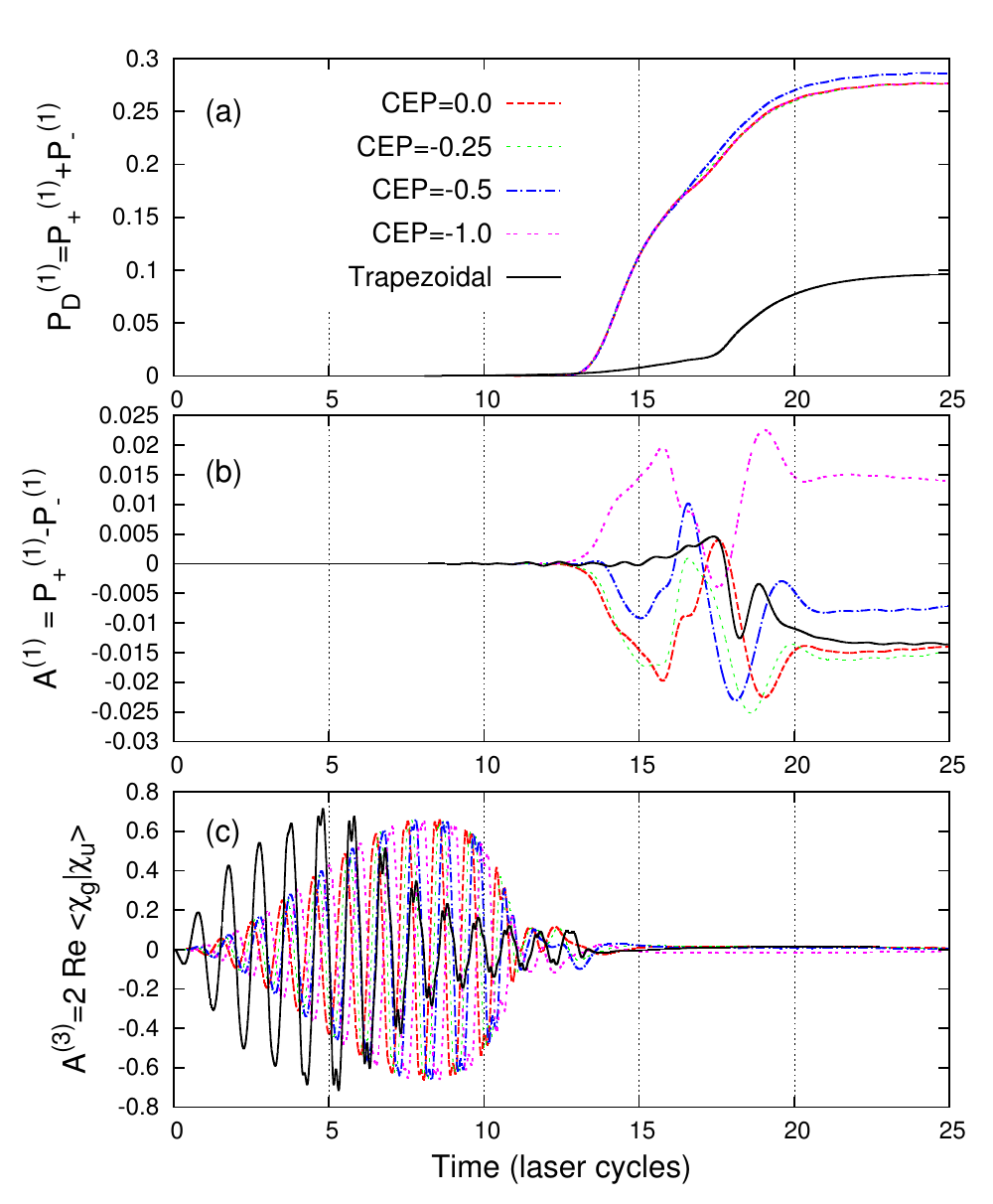}}
\caption{
(Color online) Total dissociation population $P_D^{(1)}=P_+^{(1)}+P_-^{(1)}$ (a) and  absolute asymmetry parameters $A^{(1)}$ (b) and $A^{(3)}$ (c) for corresponding spectra in Fig. 1.
}
\end{figure}
\begin{figure}[ht]
\begin{center}
\begin{tabular}{l}
\resizebox{85mm}{130mm}{\includegraphics{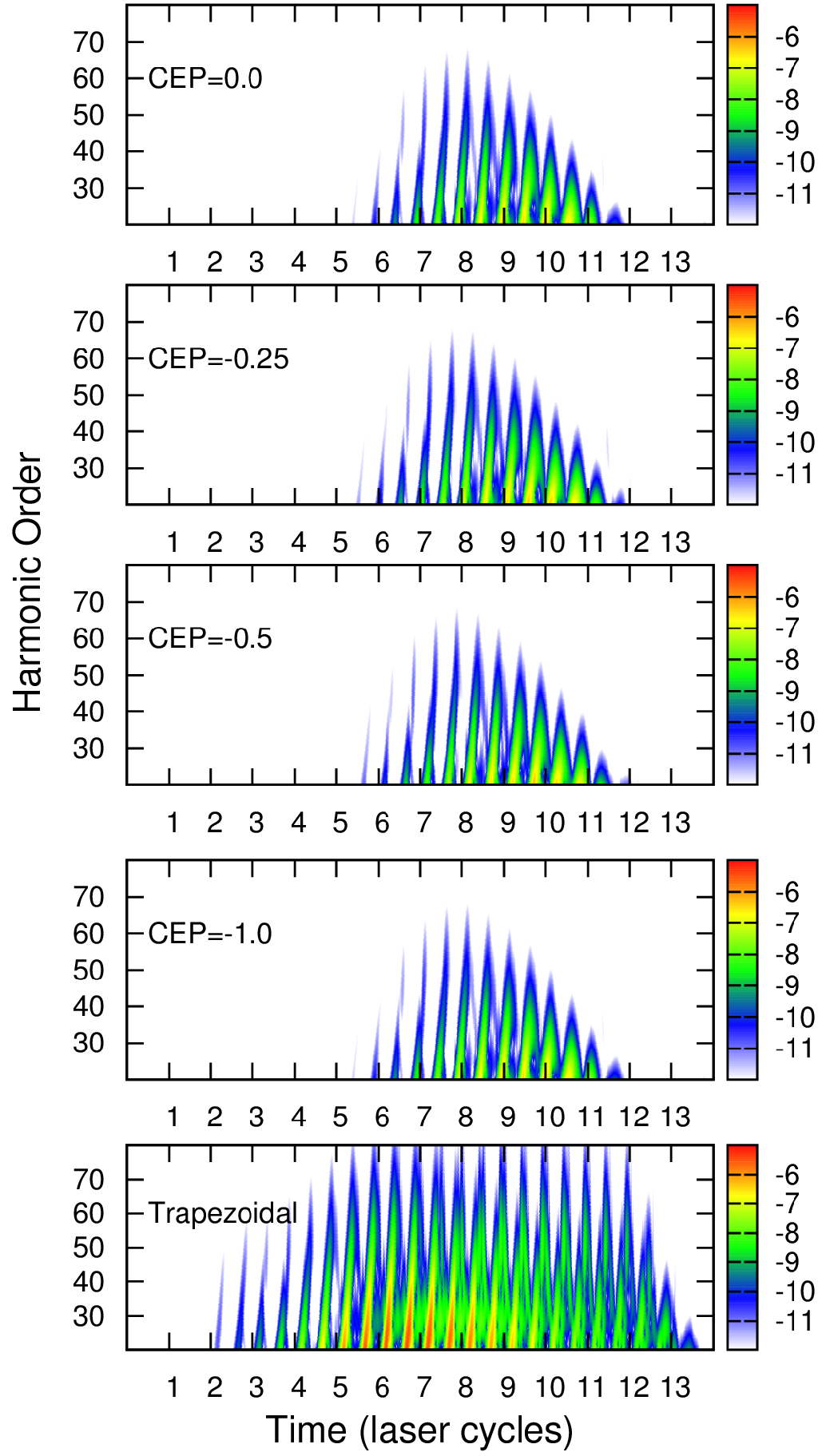}}
\end{tabular}
\caption{
\label{<z>}
(Color online) The Morlet-wavelet time profiles for NBO H$_2^+$ under 14-cycle laser pulses with sin$^2$ and trapezoidal envelopes of 800 nm wavelength and $I$=3 $\times 10^{14}$ Wcm$^{-2}$ intensity. Different CEP values are shown for the sin$^2$ case. The HHG intensities are depicted in color logarithmic scales on the right side of panels.		
		}
\end{center}
\end{figure}

\begin{figure}[ht]
\centering
\resizebox{85mm}{90mm}{\includegraphics{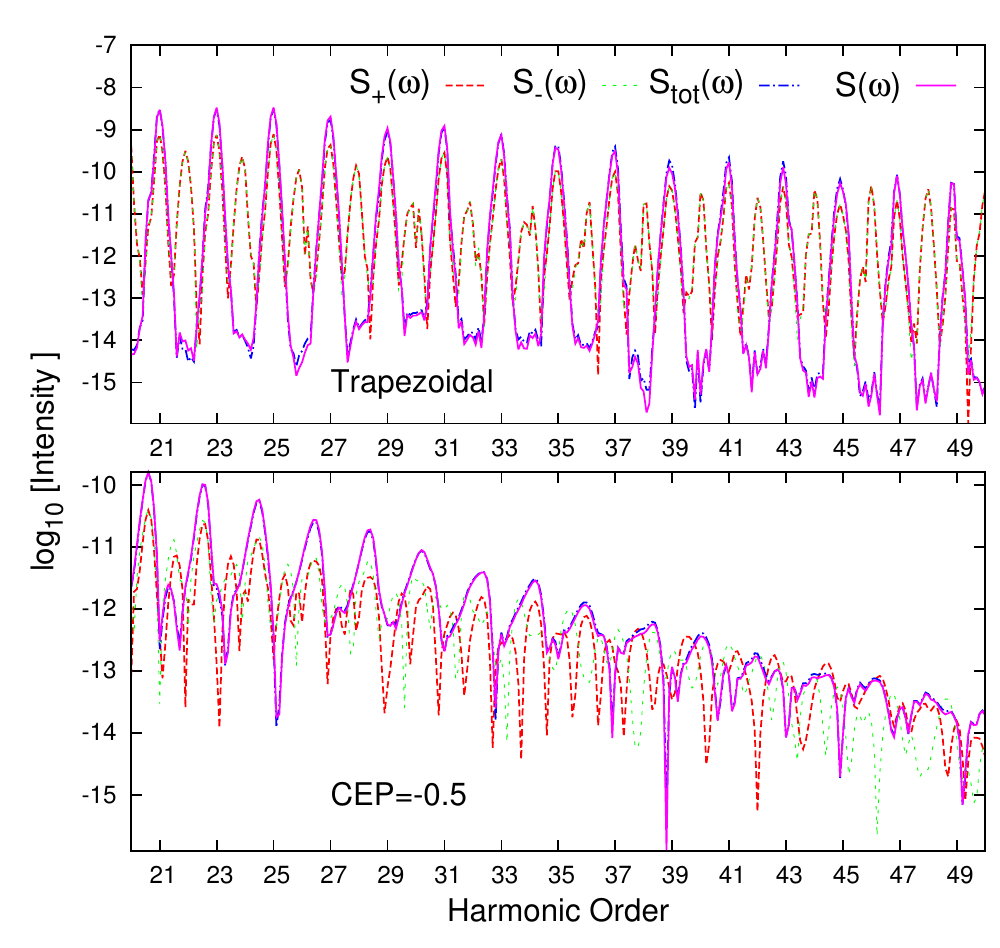}}
\caption{
(Color online) $S(\omega)$, $S_{tot}(\omega)$, $S_+(\omega)$ and $S_-(\omega)$ for two spectra of Fig. 1, trapezoidal envelope (top panel) and sin$^2$ envelope with CEP=-0.5 (bottom panel). 
}
\end{figure}

\begin{figure}[ht]
\begin{center}
\begin{tabular}{l}
\resizebox{80mm}{120mm}{\includegraphics{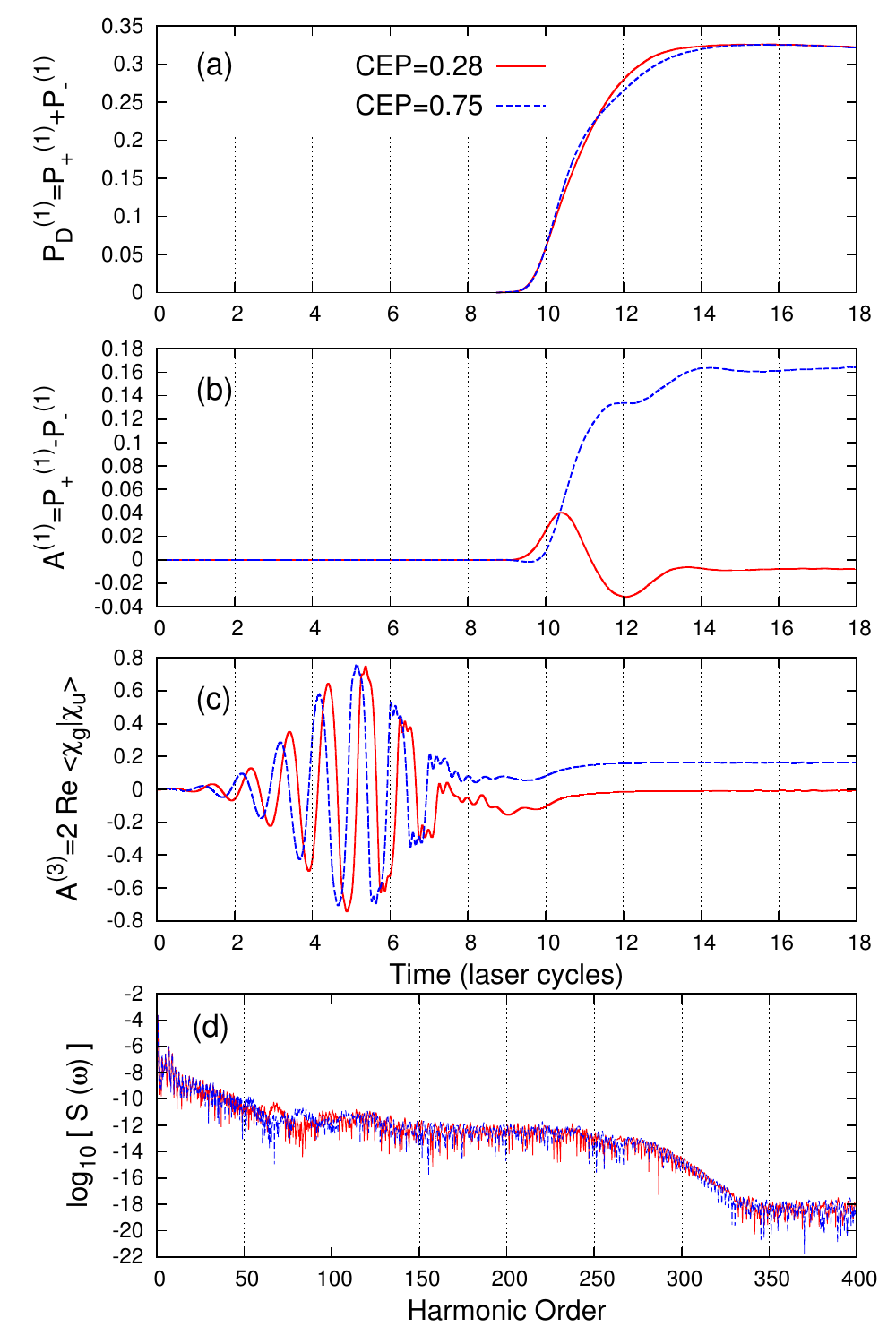}}
\end{tabular}
\caption{
\label{<z>}
(Color online) Total dissociation population $P_D^{(1)}=P_+^{(1)}+P_-^{(1)}$ (a),  absolute asymmetry parameters $A^{(1)}$ (b) and $A^{(3)}$ (c) and the HHG spectra (d). The calculations are done for 10 fs (FWHM) Gaussian laser pulses of 790 nm wavelength and $I$=9 $\times 10^{14}$ Wcm$^{-2}$ intensity with two CEP=0.28 (red line) and 0.75 (dashed blue line).		
		}
\end{center}
\end{figure}
\begin{figure}[ht]
\centering
\resizebox{85mm}{90mm}{\includegraphics{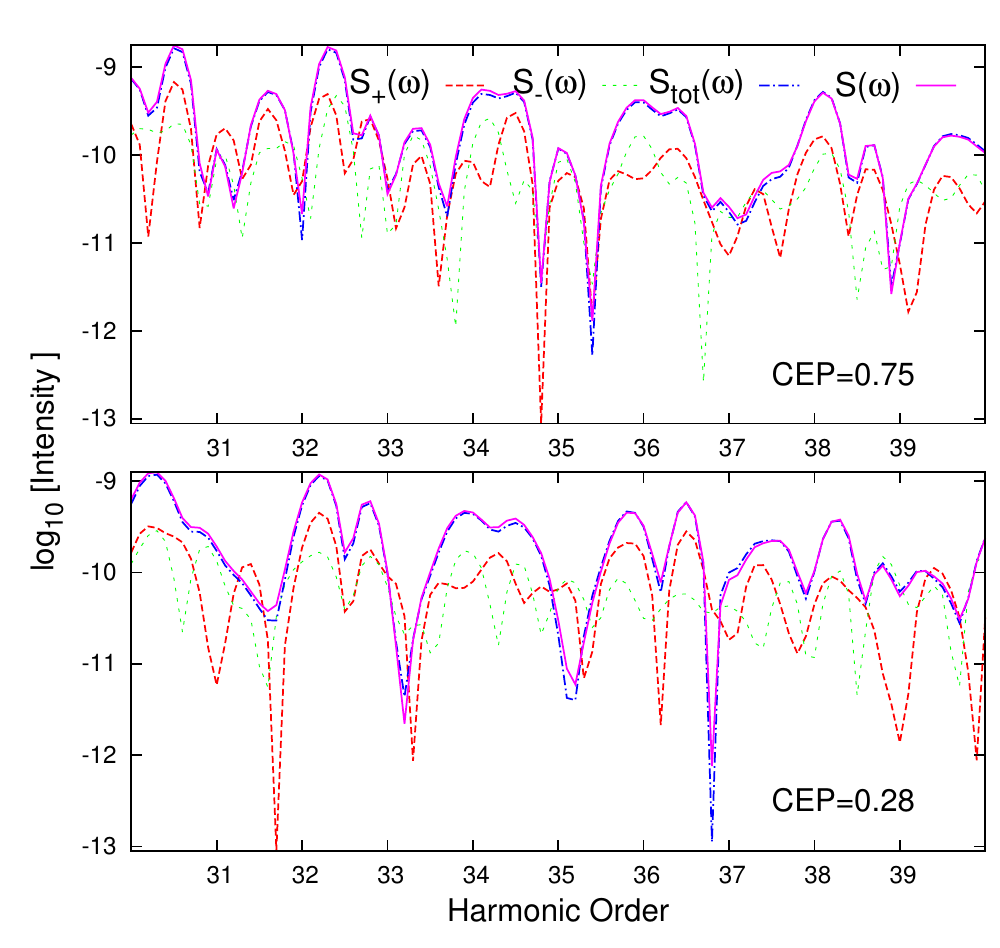}}
\caption{
(Color online) $S(\omega)$, $S_{tot}(\omega)$, $S_+(\omega)$ and $S_-(\omega)$ for H$_2^+$ in a 10 fs Gaussian laser pulses of 790 nm wavelength at $I=$9 $\times 10^{14}$ Wcm$^{-2}$ intensity, for two CEP=0.75 (top panel) and 0.28 (bottom panel). 
}
\end{figure}

The TDSE is solved using unitary split-operator methods [35-36] with a 11-point finite difference scheme through a general nonlinear coordinate transformation for both electronic and nuclear coordinates, which is described in more details in our previous works [37-39]. The grid points for $z$, $\rho$, and $R$ coordinates are 450, 100, and 300, respectively. The finest grid size values in this
adaptive grid schemes are 0.13, 0.2, and 0.025, respectively for $z$, $\rho$, and $R$ coordinates. The grids extend up to $z_{max}
= 63$, $\rho_{max} = 15$, and $R_{max}= 32$. 
The HHG spectra are calculated as the square of the windowed Fourier transform of the dipole acceleration $a_z(t)$ along the laser polarization direction (z) as
\begin{eqnarray}\label{eq:4}
  S(\omega)=\mathlarger{\vert}\int_0^T a_z(t)\,H(t)\,exp[-i\omega t]\,dt\; \mathlarger{\vert} ^2,
\end{eqnarray}
where
\begin{eqnarray}\label{eq:5}
  H(t)= \frac{1}{2}[1-cos(2\pi \frac{t}{T})],
\end{eqnarray}
is the Hanning function and $T$ is the total pulse duration. The Hanning function reduces unphysical features on the HHG spectrum as the Fourier transform is applied over a finite time by artificially cutting the dipole acceleration in Eq. (4).
The time profile of the harmonics is obtained by the Morlet-wavelet transform of the dipole acceleration $a_z(t)$ via [40-41]
\begin{eqnarray}\label{eq:6}
  &\mathlarger{w(\omega,t)= \sqrt{ \frac{\omega}{\pi^\frac{1}{2}\sigma}}\times}
 \nonumber \\
 &\mathlarger{\int_{-\infty}^{+\infty}a_z(t^\prime)exp[-i\omega (t^\prime-t)]exp[-\frac{\omega^2 (t^\prime-t)^2}{2\sigma^2}]dt^\prime.}
\end{eqnarray}
We set $\sigma=2\pi$ in this work.

 To quantify the degree of electron localization on
the two nuclei, the asymmetry parameter $A$ is defined and obtained in three ways as follows. The first absolute asymmetry parameter $A^{(1)}=P_+^{(1)}-P_-^{(1)}$ is defined with
\begin{eqnarray}\label{eq:7}
  P_{+}^{(1)}(t)=\int_{0}^{\rho_{max}}d\rho\int_{10}^{R_{max}}dR\int_{0}^{z_{max}}dz|\psi(z,\rho,R;t)|^2,   \nonumber \\
  P_{-}^{(1)}(t)=\int_{0}^{\rho_{max}}d\rho\int_{10}^{R_{max}}dR\int_{-z_{max}}^{0}dz|\psi(z,\rho,R;t)|^2 \nonumber \\   
\end{eqnarray}
in which $P_{+(-)}$ denotes to  the population on the right(left)-hand side of the simulation box in the $z$ direction for $R>10.0$ [42]. The potential curve of the first excited state of H$_2^+$, $2p\sigma_u$, flattens visibly near $R=10.0$,  and therefore we consider a molecule entering the region $R>10.0$ as the dissociative molecule. We also define $P_D^{(1)}=P_+^{(1)}+P_-^{(1)}$ as the total dissociation population which is a joint probability of finding the electron in $|z|<z_{max}$ and the protons in $10<R<R_{max}$.
As a second definition, we define $A^{(2)}=P_+^{(2)}-P_-^{(2)}$ with [43-45]
\begin{eqnarray}\label{eq:8}
  P_{\pm}^{(2)}(t)=\int_{0}^{\rho_{max}}d\rho\int_{10}^{R_{max}}dR\int_{\pm R/2.0-5.0}^{\pm R/2.0+5.0}dz|\psi(z,\rho,R;t)|^2. \nonumber \\ 
\end{eqnarray}

We found that both Eqs. (7) and (8) lead to the same final result. 
As the electron $z$ interval is considered in the vicinity of the nuclei in Eq. (8), mainly  the  $1s\sigma_g$ and $2p\sigma_u$ states contribute to the $P_{\pm}^{(2)}(t)$ and since higher excited states are less likely populated  than the $1s\sigma_g$ and $2p\sigma_u$ states, it is reasonable that $P_{\pm}^{(1)}(t)$  and $P_{\pm}^{(2)}(t)$ in Eqs. 7 and 8 give rise to the same value. 
The dissociation channel which arises from the ionization (Coulomb-explosion channel) is not important here for us, since it gives rise to no asymmetry for the protons dissociating in the opposite directions. 

To introduce the third definition of the absolute asymmetry parameter, we first decompose the total wavefunction as [24,46]
\begin{eqnarray}\label{eq:9}
  \psi(z,\rho,R;t)=& \\
   c_g(R;t)\psi_g(z,\rho;R)&+c_u(R;t)\psi_u(z,\rho;R)+\psi_{res}(z,\rho,R;t). \nonumber
\end{eqnarray}
$\psi_{g}(z,\rho;R)$ and $\psi_{u}(z,\rho;R)$ are ground and first excited electronic wavefunctions, respectively, corresponding to the $1s\sigma _g$ and $2p\sigma_u$ states. The functions $c_g(R;t)$ and $c_u(R;t)$ describe the nuclear wavepacket on the two $1s\sigma _g$ and $2p\sigma_u$ states, respectively.  The wavepacket  $\psi_{res}(z,\rho,R;t)$ is the residual part of the $\psi(z,\rho,R;t)$,  including higher excited states and electronic continuum states.
 We can also express Eq. (9) as 
\begin{eqnarray}\label{eq:10}
  \psi(z,\rho,R;t)=& \nonumber \\
   a\psi_+(z,\rho;R)&+b\psi_-(z,\rho;R)+\psi_{res}(z,\rho,R;t).
\end{eqnarray}
with
\begin{eqnarray}\label{eq:11}
  \psi_{\pm}(z,\rho;R)=1/\sqrt{2}(\psi_g(z,\rho;R)\pm\psi_u(z,\rho;R)),
\end{eqnarray}
   
\begin{eqnarray}\label{eq:12}
  a=\sqrt{2}/2(c_g(R;t)+c_u(R;t)),
\end{eqnarray}  
\begin{eqnarray}\label{eq:13}
  b=\sqrt{2}/2(c_g(R;t)-c_u(R;t)).
\end{eqnarray} 

In these equations, $\psi_{+}(z,\rho;R)(\psi_{-}(z,\rho;R))$ is the electronic wavefunction localized on the right (left) nucleus. We can define $P_+^{(3)}=|a|^2$, $P_-^{(3)}=|b|^2$ and $A^{(3)}=P_+^{(3)}-P_-^{(3)}=2Re<c_g(R;t)\vert c_u(R;t>$ [47]. It should be noted that we used the absolute asymmetry parameter instead of a normalized asymmetry parameter $A=(P_+-P_-)/(P_++P_-$),  since a small dissociation probability may lead to a large normalized asymmetry parameter.

If we substitute Eq. 10 to Eq. 4 and retain dominant terms, we arrive at 
\begin{eqnarray}\label{eq:14}
  S_{tot}\simeq S_+(\omega)+S_-(\omega)+2[A_+^*(\omega)\times A_-(\omega)],
\end{eqnarray}
where $S_+(\omega)=|A_+(\omega)|^2$ and $S_-(\omega)=|A_-(\omega)|^2$ and
\begin{eqnarray}\label{eq:15}
A_+(\omega)=& \\
  \int 2Re< a\psi_+&(z,\rho;R)\mid a_z(t)\mid \psi_{res}(z,\rho,R;t)> e^{-i\omega t} dt,\nonumber
   \end{eqnarray}
   \begin{eqnarray}\label{eq:16}
    A_-(\omega)=& \\ \nonumber
    \int 2Re< b\psi_-&(z,\rho;R)\mid a_z(t)\mid \psi_{res}(z,\rho,R;t)> e^{-i\omega t} dt. \nonumber
    \nonumber
 \end{eqnarray}
 $S_+(\omega)$ and $S_-(\omega)$ denote the recombination to the $\psi_+(z,\rho;R)$ and $\psi_-(z,\rho;R)$ states, respectively and the term $2[A_+^*(\omega)\times A_-(\omega)]$ corresponds to the electronic interference term  between these two localized electronic states.

\section{Results and Discussion}
The HHG spectra of H$_2^+$ obtained under the 14-cycle sin$^2$ (with the different CEP values) and  trapezoidal laser pulses of $I=3\times 10^{14}$  Wcm$^{-2}$ intensity and 800 nm wavelength are shown in Fig. 1. For better visualization, harmonic orders between 1-29 and 31-59 are shown separately. 
In Fig. 1, odd-order harmonics are dominant for the trapezoidal pulse while for the sin$^2$ pulses with different CEP values, odd harmonic orders are seen for low harmonics orders, which are red-shifted and broadened with increasing harmonic order. For high-order harmonics (right panel of Fig. 1), both odd and even harmonic orders are observed. 
 Morales \textit{et. al} obtained the HHG spectrum for a 1D H$_2^+$ under a 800 nm, 14-cycle sin$^2$ laser pulse (CEP=-0.5) for an intensity of $3\times 10^{14}$  Wcm$^{-2}$ [23].  They claimed that the appearance of even-order harmonics are due to a field-induced electron localization which breaks the spatial symmetry of the medium. 
We show that the observation of even harmonic orders for the sin$^2$ pulses is as a result of induced effects of the falling part of the laser pulse as opposed to what is claimed in Ref. [23].
We showed recently that even a two-cycle falling part of a trapezoidal laser pulse leads to a significant modulation on the HHG spectrum and a violation of  the odd harmonic rule [24]. 

In order to show that electron localization does not occur significantly, we have depicted $P_D^{(1)}$, $A^{(1)}$ and $A^{(3)}$ in Fig. 2, which correspond to the curves in Fig. 1. It is seen in Fig. 2(a) that the dissociation probability  increases more for the sin$^2$ pulses than the trapezoidal case. For the trapezoidal pulse, ionization is dominant over the dissociation due to a more effective amplitude of the laser pulse experienced by the molecule in the 10-cycle middle plateau. 
Regarding ionization, it is better to categorize and distinguish two different types of the ionized
electrons. In the first category, the ionized electrons are driven back to the core by the laser field and might undergo a recombination process. In the second category, the ionized electrons do not return to the core. For instance, based on the three-step model [2], electrons ionized before the peak of the laser pulse are  never driven back to the core, but those released after the pulse peak can be driven back to the core by the driving laser field. We used an absorbing
potential at the boundaries of the simulation box to avoid the second-type electron reflections
from the boundaries, leading to a decrease in the system's norm. For the trapezoidal case
with a middle plateau of 10 optical cycles, it is reasonable that the population decreases very much due
to the second-type ionization. The total norm (not shown here) at the end of the calculations decreases to 0.97 and 0.19 for the corresponding sin$^2$ cases and the trapezoidal one in Fig. 2(a), respectively. This trend is also compatible with the lower dissociation probability 
 for the trapezoidal pulse (Fig. 2(a)). The more second-type ionization occurs, the less dissociation
(based on Eqs. (7) and (8)) is observed. Therefore, a small fraction of the molecules can survive to pass through the dissociation channels through $1s\sigma_g$ and $2p\sigma_u$ states.
In Fig. 2(b), the asymmetry parameter $A^{(1)}$ is shown for the corresponding curves in Fig. 2(a). It is observed that the $A^{(1)}$ parameter goes to  small values at the end of the calculations, which indicates that  electron localization is negligible. Fig. 2(c) also shows similar behavior as in Fig. 2(b), which demonstrates a small portion of electron localization. Furthermore, the final $A^{(1)}$ value in Fig. 2(b) for the trapezoidal case (black curve) is larger than that of the sin$^2$ case  with CEP=-0.5 (dash-dotted blue curve). But, we see only odd harmonic orders for the trapezoidal pulse and even harmonic orders for the sin$^2$ one. 
 Therefore, the even harmonics seen in Fig. 1 for the sin$^2$ pulses can not be induced from  electron localization. In principle, since electron localization breaks down the symmetry of a medium, we expect to see even harmonics
throughout a whole HHG spectrum. But, if one looks carefully at the HHG spectra in Fig. 1 for harmonic orders below $\sim$ 33, these are  odd
harmonics not even harmonics which are  red-shifted, and this redshift becomes larger with
increasing harmonic order. For example, the redshift is smaller for  harmonic order 11, while it is
larger for the 27th harmonic order.

To show that induced effects of the falling part of the laser pulse is responsible for observing both even and odd harmonic orders, we plotted in Fig. 3 the corresponding Morlet-wavelet time profile  of the HHG spectra of Fig. 1. As it is obvious in this figure, for all the sin$^2$ laser pulses having different CEP values, most HHG occurs after seven optical cycles where the laser falling part is defined. Therefore, HHG gets complicated and the redshift and  complexity in the spectra increase (see [24] for more details). But for the time profile of the trapezoidal case in Fig. 3, one can see comparable HHG in the laser rising and falling  parts.

 As $P_D^{(1)}$ is smaller for the trapezoidal pulse compared to the corresponding sin$^2$ ones in Fig. 2(a), we also calculated the asymmetry parameters and the HHG spectrum for a 14-cycle trapezoidal pulse with a lower intensity of $I$=2.5$\times 10^{14}$ Wcm$^{-2}$  in order to have a considerable $P_D^{(1)}$ comparable to that of the sin$^2$ pulses shown in Fig. 2(a). We did not observe any considerable asymmetry and only  odd-order harmonics appeared for this trapezoidal laser pulse. It is less likely to observe electron localization for even longer trapezoidal pulses ($>$ 14 optical cycles) since the second-type ionization would be higher and no electron will remain for the molecule to be localized on either each of the nuclei around an intensity of 3$\times 10^{14}$ Wcm$^{-2}$. Furthermore, for the trapezoidal and sin$^2$ pulses with 14 optical cycles, there is enough time for the molecule to reach the region $R>6$ where electron localization is more probable, while as stated above, we did not observe any electron localization.

  As laser pulse duration increases, larger internuclear distances become accessible, and we can say that higher vibrational states are populated (below saturation laser intensity). These high-lying vibrational states may contribute to both even and odd harmonic orders. We showed that as long as the falling edge of a trapezoidal laser pulse is insignificant, we see only the contribution of these higher vibrational states to  odd harmonic orders [24]. But these high-lying vibrational states contribute mainly to even harmonic orders due to an effective contribution of the laser falling part [24]. Therefore, it is hard to attribute the existence of even harmonic orders to electron localization, as done in Ref. [23], without considering induced effects of the falling edge of a laser pulse. The best way one can ensure whether there is a degree of electron localization is to calculate the asymmetry parameters, as presented in this work. 
  
 The complex patterns in the HHG spectra for the sin$^2$ cases in Fig. 1 can be originated from two effects. The first observation is a non-adiabatic frequency redshift of the harmonics which is clearly seen for  low harmonic orders on the left panel of Fig. 1. This effect is almost independent of the CEP and a similar frequency redshift is seen for different CEP values in Fig. 1 (left panel).  Another effect comes from a spatially asymmetric emission along the $z$ direction which breaks down the odd-harmonic rule.  We recently showed that this asymmetric emission along the $z$ direction, which is the same direction of laser polarization direction, can occur even in the falling part of trapezoidal laser pulses depending on the pulse duration, laser intensity and type of the isotope [25]. It is shown in Fig. 3 that most harmonic emission occurs in the laser falling part. The time-dependent laser intensity decreases from one cycle to another in the falling part.  Suppose that in the laser falling part the electron releases in a laser half cycle along the negative $z$ direction and can also be driven back to the core and mainly recombine with the nucleus located along the negative direction (see Ref. [48] for a similar work). In the next successive half cycle upon the field sign reversal, the electron should be ionized along the positive-$z$ direction and similarly recombine to the other nucleus along the positive-$z$ direction.  Since the laser intensity decreases in time in the falling part, the HHG symmetry along both  negative and  positive $z$ directions breaks down, leading to a even-order-harmonic generation.
 For the harmonic orders above 37 both the non-adiabatic effect and spatial symmetry breaking are present which make the HHG spectra more complicated. For low harmonic orders $<$37, the harmonic emission is not influenced considerably by the decreasing electric field at the pulse falling part as compared to high harmonic orders ($>$ 37). 
We can also observe both the non-adiabatic effect and  the spatially asymmetric emission by the analysis of  
 $S(\omega)$, $S_+(\omega)$, $S_-(\omega)$ and $S_{tot}(\omega)$ (see Eqs. (4) and (14)) as  demonstrated in Ref. [25]. We have depicted these components in Fig. 4 for the trapezoidal envelope (top panel) and a sin$^2$ case with CEP=-0.5 (bottom panel) of Fig. 1. $S(\omega)$ and  $S_{tot}(\omega)$ are almost overlapped, demonstrating that the approximation made in Eq. (14) is satisfactory. In other words, the recombination into the $1s\sigma_g$ and $2p\sigma_u$  states are dominant in the HHG process.
  One can see that $S_+(\omega)$ and $S_-(\omega)$  for the trapezoidal pulse show both even and odd harmonic orders with a comparable intensity.  For $S_{tot}(\omega)$ one see that even harmonic orders are suppressed significantly but odd ones are intensified. The suppression of even harmonics is due to the interference term $2[A_+^*(\omega)\times A_-(\omega)]$ which can be considered as an interference term between the two localized left and right wavepackets. In contrast, for  sin$^2$ case in Fig. 4 (bottom panel) $S_+(\omega)$ and $S_-(\omega)$ show complicated patterns. Both the non-adiabatic effect and the effect of the spatially asymmetric emission on the HHG spectrum can be deduced in the sin$^2$ case. The non-adiabatic effect is purely seen at low harmonic orders as it is also obvious in the left panel of Fig. 1. For example, for the peak corresponding to the 21st harmonic order we see both $S_+(\omega)$ and $S_-(\omega)$ are comparably overlapped. This redshifted peak (the peak in Fig. 4 on the bottom panel between harmonic order 20 and 21) is as a result of the non-adiabatic effect. But for higher harmonic orders, $S_+(\omega)$ and $S_-(\omega)$ lose their overlap which we attribute it to the spatially asymmetric emission along the positive and negative $z$ directions.       
   
In order to have a substantial final electron localization, we calculated the  asymmetry parameters and the HHG spectrum for a 10 femtosecond (full width at half maximum) Gaussian laser pulse of 790 nm wavelength and $I$=9$\times 10^{14}$ Wcm$^{-2}$ intensity for the two CEP values of 0.75 and 0.28, which are shown in Fig.  5 (all the calculation and laser parameters are chosen based on Ref. [43]). In Fig. 5(a), we see a significant value for $P_D^{(1)}$, which is essential to probably see a considerable asymmetry. We see in Fig. 5(b) that $A^{(1)}$  is considerably higher for the CEP=0.75  than the CEP=0.28. Therefore, we can say that electron localization has happened for the CEP=0.75 as it can be also deduced from the $A^{(3)}$ parameter in Fig. 5(c). The HHG spectra for both CEP values are shown in Fig. 5(d) in which we observe both odd and even harmonic orders for both CEP values. Figure 6 also shows $S_+(\omega)$, $S_-(\omega)$, and $S_{tot}(\omega)$, corresponding to the HHG spectra in Fig. 5(d).  $S_+(\omega)$ and $S_-(\omega)$ do not overlap generally even for the CEP=0.28 case with the negligible electron localization. That also rationalizes the occurrence of the spatially asymmetric emission along the $z$ direction.

\section{Conclusion}
We solved numerically the full-dimensional electronic time-dependent Schr\"{o}dinger equation for H$_2^+$ beyond the Born-Oppenheimer approximation to resolve complex patterns observed in  high-order harmonic generation under intense sin$^2$-like laser pulses.  The contribution from  electron localization, non-adiabatic redshift and spatially asymmetric emission was demonstrated to understand better the complexities. We considered long laser pulses with Gaussian, sin$^2$ and trapezoidal laser envelopes to investigate the effect of electron localization on the HHG process. For the trapezoidal laser pulse, no considerable  electron localization was achieved and the HHG spectrum was dominant by odd harmonic orders. For 14-cycle sin$^2$ laser pulses with $I$=3$\times 10^{14}$ Wcm$^{-2}$ intensity and for different CEP values, no significant final electron localization was found. We observed that most HHG process occur at the falling part of the laser pulse due to the nuclear motion. We showed that complicated patterns appear due to two effects.
The first effect originates from the non-adiabatic response of the molecule to the rapidly changing laser field in the laser falling part. This effect is more visible at low harmonic orders, which results in a frequency redshift of the harmonics. The second effect comes from the spatially asymmetric emission along the polarization direction at the laser falling part. Both effects make the HHG spectrum complex for high harmonic orders. We also decomposed the total harmonic signal into different localized signals so that we are able to deduce both effects in the HHG process.
It was shown that the appearance of even harmonic orders is not a good criterion to conclude that final electron localization has happened due to the significant effects of a falling edge of sin$^2$-like laser pulses.

\FloatBarrier 
\section{References}
\bibliography{p7}

\end{document}